\documentclass[aps,english,twocolumn,prb,superscriptaddress,showpacs]{revtex4-1}
\usepackage{units}
\usepackage{amsmath}
\usepackage{amssymb}
\usepackage{graphicx}
\usepackage{url}

\usepackage{babel}
\begin{document}

\title{Quantum order by disorder in the Kitaev model on a triangular lattice}

\author{George Jackeli}

\altaffiliation{Also at E. Andronikashvili Institute of Physics, 0177 Tbilisi, Georgia}

\selectlanguage{english}%

\affiliation{Institute for Functional Matter and Quantum Technologies, University
of Stuttgart, Pfaffenwaldring 57, D-70569 Stuttgart, Germany}

\email{G.Jackeli@fkf.mpg.de}

\selectlanguage{english}%

\affiliation{Max Planck Institute for Solid State Research, Heisenbergstrasse
1, D-70569 Stuttgart, Germany}

\author{Adolfo Avella}

\affiliation{Dipartimento di Fisica ``E.R. Caianiello'', Universit\`a degli Studi
di Salerno, I-84084 Fisciano (SA), Italy}

\email{avella@physics.unisa.it}

\selectlanguage{english}%

\affiliation{CNR-SPIN, UoS di Salerno, I-84084 Fisciano (SA), Italy}

\affiliation{Unit\`a CNISM di Salerno, Universit\`a degli Studi di Salerno, I-84084
Fisciano (SA), Italy}
\begin{abstract}
We identify and discuss the ground state of a quantum magnet on a triangular lattice with bond-dependent Ising-type spin couplings, that is, a triangular analog of the Kitaev honeycomb model. The classical ground-state manifold of the model is spanned by decoupled Ising-type chains, and its accidental degeneracy is due to the frustrated nature of the anisotropic spin couplings. We show how this subextensive degeneracy is lifted by a quantum order-by-disorder mechanism and study the quantum selection of the ground state by treating short-wavelength fluctuations within the linked cluster expansion and by using the complementary spin-wave theory. We find that quantum fluctuations couple next-nearest-neighbor chains through an emergent four-spin interaction, while nearest-neighbor chains remain decoupled. The remaining discrete degeneracy of the ground state is shown to be protected by a hidden symmetry of the model. 
\end{abstract}

\pacs{75.10.Jm, 75.30.Ds, 75.30.Et}

\date{\today}

\maketitle
Frustrated magnets, systems in which every pairwise exchange interaction
cannot be simultaneously satisfied, are characterized by accidental
degeneracies between various order patterns \cite{Bal10}. Often,
these accidental degeneracies are lifted via an order-by-disorder
mechanism, driven by thermal and/or quantum fluctuations, selecting
an unique ground state \cite{Vil80,She82, [{For the discussions on experimental manifestation of quantum order-by-disorder see }]Zhi12,*Sav12}. In highly frustrated
quantum magnets, those with extensive degeneracy, e.g., the isotropic
spin one-half kagom\'e and pyrochlore antiferromagnets (AF), the order-by-disorder
mechanism is inactive and they remain disordered down to the lowest
temperatures, realizing so-called quantum spin liquids (QSL) in their
ground states \cite{Bal10}.

In Mott insulators, with unquenched orbital moments and strong spin-orbit
coupling, bond-dependent Ising-type interactions may dominate over
the conventional Heisenberg term \cite{Kha05,Che08,Jac09}. In turn,
such Ising-type couplings, even being ferromagnetic (FM), can frustrate a
long-range magnetic order  and stabilize a QSL state \cite{[{For a review see }]Nus13,*Nus15}. The most celebrated
model realizing the above scenario is the exactly solvable Kitaev
honeycomb model \cite{Kit06}. In this model, nearest-neighbor (NN)
spins are coupled by Ising-type terms and the three non-equivalent
bonds of the honeycomb lattice host different components of the spin
one-half operators. Its ground state is a QSL with fractionalized
fermionic excitations \cite{Kit06}.

Following a theoretical proposal \cite{Jac09} for a possible realization
of the Kitaev honeycomb model in iridates \textit{A}$_{2}$IrO$_{3}$
(\textit{A} = Na, Li), various extensions of the model have been studied
in connection to experiments \cite{Sin10,Liu11,Cho12,Sin12,Ye12,Gre13,Chu15}
on actual materials. These model extensions include terms like: the
isotropic Heisenberg exchange (the so-called Kitaev-Heisenberg (KH)
model) \cite{Cha10,Jia11,Reu11,Pri12,Cha13}, further-neighbor couplings
\cite{Kim11,Cha13,Reu14}, and additional symmetry-allowed anisotropies
\cite{Kat14,Rau14,Yam14,Siz14,Cha15}. The resulting theoretical phase
diagrams are characterized by various ordered phases (including those
seen experimentally) and by a finite stability window for QSL around
the Kitaev limit.

Recently, a triangular analog of the KH model \footnote{The KH triangular model is an extention of the anisotropic spin model
originally proposed and studied in Ref.~\onlinecite{Kha05} in
the context of sodium cobaltates.} for classical \cite{Rou12} and quantum \cite{Bec14,Li14} spins
has been studied numerically. The obtained rich phase diagram includes
a $\mathbb{Z}_{2}$-vortex crystal phase near the AF Heisenberg limit, and a nematic phase of decoupled Ising chains with
sub-extensive degeneracy at the Kitaev limit \cite{Rou12,Bec14}. In addition,
a chiral spin-liquid phase has been proposed close to the antiferromagnetic
Kitaev limit \cite{Li14}.

Here, we study analytically  the Kitaev model on the triangular lattice and solve
the puzzle of its ground state by analyzing the effects of quantum
fluctuations within both the linked-cluster expansion, \cite{Gel00}
combined with degenerate perturbation theory, and the linear spin-wave
theory. We show that such a deceptively simple model, once realized
on a triangular lattice, becomes the host of very interesting and
unexpected order-by-disorder effects such as: the quantum selection
of the easy axes, the emergence of a specific four-spin
interaction, the reduction of the sub-extensive degeneracy of the
\emph{nematic} ground state manifold down to a discrete one protected
by a hidden symmetry of the model.

\section{The model}

\begin{figure}[tb]
\centering{}\includegraphics[width=8cm]{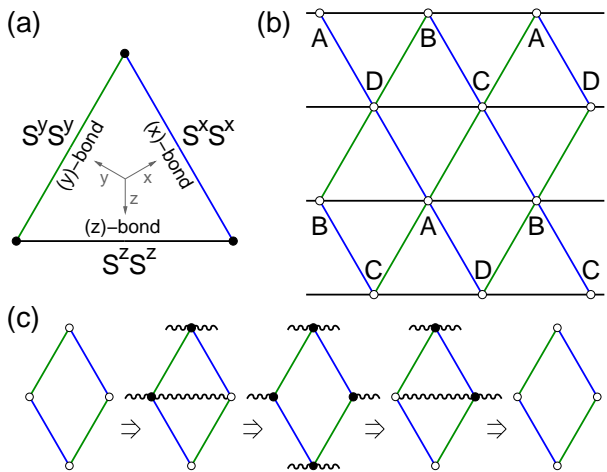}\protect\caption{(color online). (a) Ising-type spin couplings on the three non-equivalent
bonds of the triangular lattice of model (\ref{eq1}). The lattice
lies in the $(1,1,1)$ plane of the spin-quantization axes. On a $(\gamma)$-bond,
the one perpendicular to the $\gamma\:(=x,\:y,\:z)$ spin quantization
axis, only the $\gamma$-components of neighboring spins are coupled.
(b) Four-sublattice structure of the triangular lattice used for the
unitary transformations discussed in the text. (c) Sketch of the 4$^{th}$
order perturbation process leading to the coupling, via quantum fluctuations,
of the four spins siting around a \emph{diamond} {[}see Eq.~(\ref{eq3}){]}.
In the virtual states, the location of the misaligned spins (spin-flips
are performed in pairs at one of the 4 bonds of the \emph{diamond}
at each step: top-left $\Rightarrow$ bottom-right $\Rightarrow$
bottom-left $\Rightarrow$ top-right) are shown by filled circles
and the wavy lines mark the \emph{broken} $(z)$-bonds.\label{fig1}}
\end{figure}

We consider a triangular lattice lying in the
$(1,1,1)$ plane of the spin-quantization frame {[}see Fig.~\ref{fig1}(a){]}
and label by $(\gamma)\:(=x,\:y,\:z)$ its three non-equivalent NN
bonds spanned by the lattice vectors $\mathbf{a}_{x}=\left(\nicefrac{1}{2},-\nicefrac{\sqrt{3}}{2}\right)$,
$\mathbf{a}_{y}=\left(\nicefrac{1}{2},\nicefrac{\sqrt{3}}{2}\right)$
and $\mathbf{a}_{z}=\left(1,0\right)$, respectively. On a $(\gamma)$-bond,
the one perpendicular to the $\gamma$ spin-quantization axis, only
the $S_{\mathbf{i}}^{\gamma}$ components of the spin one-half operators
$\mathbf{S}_{\mathbf{i}}$ are coupled by a Ising-type interaction
{[}see Fig.~\ref{fig1}(a){]}, and the corresponding Hamiltonian
takes the following form

\begin{equation}
{\cal H}=-\sum_{\mathbf{i},\gamma}K_{\gamma}S_{\mathbf{i}}^{\gamma}S_{\mathbf{i}+\mathbf{a}_{\gamma}}^{\gamma}\,.\label{eq1}
\end{equation}

In model (\ref{eq1}), the signs of the $K_{\gamma}$ couplings can
be individually flipped by means of a canonical transformation. For
instance, to flip the sign of $K_{z}$ independently from the signs
of the other two couplings, $K_{x}$ and $K_{y}$, one needs to perform
spin rotations around the $y$ axis by an angle $180^{\circ}$ on
sites belonging to the sublattices $\mathsf{B}$ and $\mathsf{C}$
{[}see Fig.~\ref{fig1}(b){]}, i.e., {[}$\left(S_{\mathbf{i}}^{x},S_{\mathbf{i}}^{y},S_{\mathbf{i}}^{z}\right)\rightarrow\left(-S_{\mathbf{i}}^{x},S_{\mathbf{i}}^{y},-S_{\mathbf{i}}^{z}\right)$
for $\mathbf{i}\in\mathsf{B}\oplus\mathsf{C}${]}. The signs of $K_{x}$
($K_{y}$) can be flipped independently in the very same way by performing
$180^{\circ}$ spin rotations around $z$ ($x$) axis on the sublattices
$\mathsf{B}$ and $\mathsf{D}$ ($\mathsf{A}$ and $\mathsf{B}$).
In what follows, without any loss of generality, we consider all $K_{\gamma}$
to be positive (FM couplings).

\section{Ground state manifold}

In the isotropic FM case $K_{\gamma}=K>0$,
the classical ground-state energy is simply proportional to $\mathbf{M}^{2}$
where $\mathbf{M}=\left\langle \mathbf{S}_{\mathbf{i}}\right\rangle $.
This accidental symmetry implies that the ordered moment ${\bf M}$
can be freely rotated, i.e., no preferred axis exists. Moreover, the
coupling between NN chains, along any of the three lattice directions
(e.g., spanning along $(z)$-bonds), does not involve the corresponding
projections of the spins (e.g., $S_{\mathbf{i}}^{z}$). Therefore,
these latter projections of the spins can be freely flipped along
any of those chains individually \cite{Rou12,Bec14}. This leads to
an additional $2^L$-fold degeneracy, where $L$ is the linear size of the system.
In the anisotropic case, when the
couplings $K_{\gamma}$ are different from one another, the easy axis
is dictated by the strongest coupling (e.g., $z$ axis for $\left|K_{z}\right|>\left|K_{x}\right|,\left|K_{y}\right|$).
However, the ground-state manifold still has a sub-extensive degeneracy
as it is characterized by completely decoupled either FM (for $K_{z}>0$)
or AF (for $K_{z}<0$) chains along $(z)$-bonds. Such a sub-extensive degeneracy
is inherent to models with Ising- or compass-type bond-dependent anisotropies 
\cite{Nus13,*Nus15}.

In principle, these accidental classical degeneracies, not being related
to apparent symmetries, can be lifted by quantum fluctuations. We
would need to calculate the energy corrections due to zero-point quantum
fluctuations (e.g., within the spin-wave theory) for each degenerate
classical ground state and single out a ground state for which the
corrected energy is minimized. For an infinitely degenerate manifold this
is obviously not feasible and we need to resort to some other procedure.
The linked-cluster expansion, \cite{Gel00} combined with degenerate
perturbation theory, allows to compute quantum corrections to a ground-state
energy from short-wavelength quantum fluctuations and to identify
the mechanism for quantum selection of the ground state. \cite{Lon89,Hei93,Ber07,Mar14,Che14,Zhi15}

\section{Quantum selection of the ground state}

\subsection{Easy axes}

In the isotropic case $K_{\gamma}=1$, we consider a FM state with
the ordered moment ${\bf M}$ pointing in a generic direction identified
by the unit vector ${\bf m}=\left(m_{x},m_{y},m_{z}\right)=\left(\sin\theta\cos\phi,\sin\theta\sin\phi,\cos\theta\right)$.
Then, we rotate the spin-quantization frame $xyz$ of the Hamiltonian
(\ref{eq1}) to a new frame $x'y'z'$ in which ${\bf m}\parallel z'$.
The transformed Hamiltonian on NN $ij$ bond includes various terms
in the new spin-quantization frame: the $S_{i}^{z'}S_{j}^{z'}$ terms
represent the unperturbed (mean-field) Hamiltonian and the remaining
ones, those creating misaligned spins at the cost of a mean-field
energy, are treated as perturbations. At the second order in the perturbation
expansion, the terms creating only one spin-flip, e.g., $S_{i}^{x'}S_{j}^{z'}$,
give energy corrections that sum up to zero. Only the terms inducing
two spin-flips on a given ($\gamma$)-bond give a cumulative finite
energy correction depending on the direction of ${\bf m}$. The creation/annihilation
amplitude for two misaligned spins on a ($\gamma$)-bond is $T_{\gamma}=\left(1-m_{\gamma}^{2}\right)/4$
with a corresponding energy cost $\Delta_{\gamma}=(2-m_{\gamma}^{2})$.
This gives the following quantum energy correction
per site:

\begin{equation}
\delta E^{(2)}({\bf m})=-\sum_{\gamma}\frac{T_{\gamma}^{2}}{\Delta_{\gamma}}\simeq-\frac{3}{64}\left(1+\frac{1}{6}\sum_{\gamma}m_{\gamma}^{4}\right)\label{eq2}
\end{equation}

\begin{figure}[tb]
\centering{}\includegraphics[width=6cm]{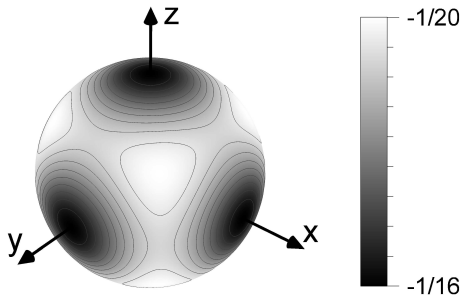}\protect\caption{Color map of the second-order quantum energy correction (\ref{eq2})
for an arbitrary direction ${\bf m}$ of the ordered moment. The minimum
energy is achieved for ${\bf m}$ pointing along one of the spin-quantization
axes.\label{fig2}}
\end{figure}

In Fig.~\ref{fig2}, we report the second-order quantum energy correction
(\ref{eq2}) for an arbitrary direction ${\bf m}$ of the ordered
moment via a color map. Each point on the sphere
stands for a specific direction of ${\bf m}$ within the original
spin-quantization frame $xyz$ and the color scale gives the corresponding
value of (\ref{eq2}). It is evident that the minimum of the energy
is realized when ${\bf m}$ points along one of the spin-quantization
axes implying that they are selected by quantum corrections as the easy
axes. This can be also seen from the result of a \emph{naive} expansion,
the last term in Eq.~(\ref{eq2}), explicitly showing how a fourth-order
cubic anisotropy emerges from quantum fluctuations.

The above result explains and quantifies the order-by-disorder selection
of the easy axes numerically found in Refs.~\onlinecite{Rou12,Bec14}
by means of classical Monte Carlo and of density matrix renormalization
group (DMRG) methods, respectively, and also agrees with previous
studies performed on models similar to (\ref{eq1}), but realized
on other lattices \cite{Kha01,Cha10,Sel14,Siz15}. This study is complementary to the spin-wave analysis performed in Ref.~\onlinecite{Bec14}
around the Heisenberg limit of the FM KH model on a triangular lattice. 
In that case,  a finite Kitaev coupling leads to the selection of easy axes and breaks 
the accidental $O\left(3\right)$ symmetry down to $\mathbb{Z}_{6}$.
In the present case, the degenracy of the ground state remains sub-extensive at $3\times2^L$.

\subsection{Coupling between chains}

Both in the isotropic case $K_{\gamma}=K$
(choosing $z$ as easy axis) and in the anisotropic case $\left|K_{z}\right|\geq\left|K_{x}\right|,\left|K_{y}\right|$,
the ground-state manifold is spanned by decoupled either FM (for $K_{z}>0$)
or AF (for $K_{z}<0$) chains along $(z)$-bonds. We now compute,
within the linked-cluster expansion \cite{Gel00}, the quantum corrections
induced by the $K_{x}$ and $K_{y}$ terms, and find the related effective
couplings between chains. The expansion parameter scales as $K_{x(y)}
/{\rm z}K_z$ (${\rm z}=3$ being the number of NNs), and the results we derive are
valid in the isotropic case $K_{\gamma}=K$ too. The $K_{x}$ and $K_{y}$  terms generate fluctuations
out of the classical ground state by creating/annihilating pairs of
misaligned spins on the corresponding bonds. Accordingly, the linked
graphs relevant to the perturbation expansion are those composed by
$(x)$- and/or $(y)$- bonds. Open linked graphs composed of $n$
bonds contribute to the $\left(2n\right)^{th}$ leading order, while
closed linked graphs composed of $n$ bonds contribute to the $n^{th}$
leading order. Given the obvious absence of $(z)$-bonds in the perturbation
expansion, there is no loop-like cluster composed of an odd number
of bonds and hence no perturbative odd-order correction exist. In
the $2^{nd}$ order, there is no coupling between chains but a reduction
of the ground state energy by $-\frac{1}{32}\frac{K_{x}^{2}+K_{y}^{2}}{\left|K_{z}\right|}$.
In the $4^{th}$ order, straight 3-site clusters, composed of two
$(x)$- or two $(y)$- bonds, give just a reduction of the ground
state energy by $-\frac{1}{2048}\frac{\left(K_{x}^{2}+K_{y}^{2}\right)^{2}}{\left|K_{z}^{3}\right|}$
and again do not couple chains. The other two types of 3-site clusters,
composed of $(x)$- and $(y)$- bonds, with $\pi/3$ or $2\pi/3$
angles in between, give no contribution at all because the ${\cal H}^{\left(x\right)}$
and ${\cal H}^{\left(y\right)}$ bond Hamiltonians with a common vertex
anticommute and they always come in permuted pairs. The $4^{th}$
order correction coming from a \emph{diamond}-shape 4-site cluster
{[}see Fig.~\ref{fig1}(c){]} give instead a coupling between pairs
of spins belonging to next-nearest-neighbor (NNN) chains in the following
form

\begin{equation}
\delta{\cal H}^{(4)}=-\frac{1}{24}\frac{K_{x}^{2}K_{y}^{2}}{\left|K_{z}^{3}\right|}\sum_{\mathbf{i}}\left(S_{\mathbf{i}}^{z}S_{\mathbf{i}+\mathbf{a}_{z}}^{z}\right)S_{\mathbf{i}+\mathbf{a}_{x}}^{z}S_{\mathbf{i}+\mathbf{a}_{y}}^{z}\label{eq3}
\end{equation}

where the sites $\mathbf{i}+\mathbf{a}_{x}$ and $\mathbf{\mathbf{i}}+\mathbf{\mathbf{a}_{y}}$
belong to NNN chains, and they are the ends of the longer diagonal
of the \emph{diamond} cluster {[}see Fig.~\ref{fig1}(c){]}. It is
worth noting that $S_{\mathbf{i}}^{z}S_{\mathbf{i}+\mathbf{a}_{z}}^{z}$
is just $\frac{1}{4}$ for $K_{z}>0$ and $-\frac{1}{4}$ for $K_{z}<0$
leading to a coupling between NNN chains of the same \emph{sign} of
the one acting along the chains. This does not fully lift the degeneracy
as, at this order, the two sublattices formed by NN chains remain
decoupled. The degeneracy is four-fold (3 times four-fold for $K_{x}=K_{y}=K_{z}$).
Actually, we found that this degeneracy is dictated by a hidden symmetry
of the model. This hidden symmetry is uncovered by a four-sublattice
unitary transformation from Ref.~\onlinecite{Kha05}. This transformation
leaves the Hamiltonian (\ref{eq1}) unchanged, but flips the sign
of the $z$ components of the spins on only one of the two sublattices.
We divide the triangular lattice in 4 sublattices, as shown in Fig.~\ref{fig1}(b),
and perform the following local spin-rotations: by $180^{\circ}$
on sublattices $\mathsf{B}$, $\mathsf{C}$, and $\mathsf{D}$ around
$z$, $y$, and $x$ axes, respectively, while keeping the spins on
the $\mathsf{A}$ sublattice in the original frame. This transformation
leaves the Hamiltonian (\ref{eq1}) unchanged. On the other hand,
the net effect on a state with spins ordered along the $z$ direction
is to flip the sign of the $z$ component of every second chain, the
chains belonging to the $\mathsf{C}\oplus\mathsf{D}$ sublattice,
showing that NN chains are completely decoupled as any relative order
between them leads to the same energy.

\subsection{Comparison to numerics}

Very recently, Becker \textit{et.
al}, \cite{Bec14} used the DMRG method to compute the ground state
and the first excited states of the Kitaev Hamiltonian (\ref{eq1})
in the AF isotropic case ($K_{\gamma}=K<0$) on finite clusters with
open boundary conditions. The considered clusters are strips of 3
and 4 chains of length $L\leq14$, cut out from a triangular lattice
(see Figs.~12 and 13 in Ref.~\onlinecite{Bec14}). The spatial
anisotropy of such clusters breaks the original symmetry of the triangular
lattice and forces the spins to order AF along the longer direction
(e.g., along the $(z)$-bonds) and, correspondingly, in the like spin
component ($S^{z}$). Measuring the spin correlation functions across
the system, they found AF correlations between the NNN chains and
no correlations between NN ones, in agreement with our analysis. Moreover,
for the largest analyzed system, the numerically found gap to the
first excited state, featuring FM correlations between NNN chains,
amounts to $0.0055\left|K\right|$ per \emph{diamond}, which is again
in very good agreement with the value $1/192\left|K\right|\simeq0.0052\left|K\right|$
predicted by Eq.~(\ref{eq3}).

\section{Spin-wave theory}

\begin{figure}[tb]
\centering{}\includegraphics[width=8cm]{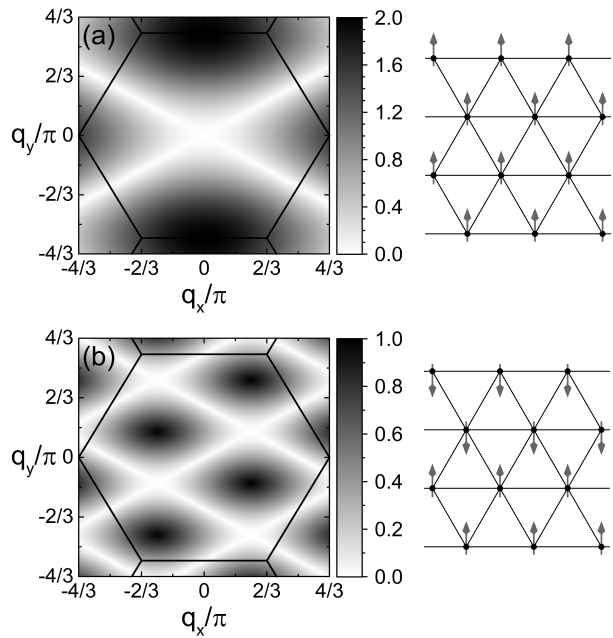}\protect\caption{Color map of the spin-wave excitation spectrum of model (\ref{eq1})
in (a) the FM state Eq.~(\ref{eq4}) and (b) the stripy AF state
Eq.~(\ref{eq5}), for $K_{\gamma}=1$ . Only one of the lower branches
is shown in (b). Hexagons mark the crystallographic Brillouin zone.
The right panels report the corresponding magnetic structures.\label{fig3}}
\end{figure}

We now apply the linear spin-wave theory
to the Hamiltonian (\ref{eq1}). In order to compare the results obtained
by the linear spin-wave theory with those obtained by the linked-cluster
expansion, we will focus on the case $K_{z}\geq K_{x},K_{y}>0$ and
consider two states that are degenerate in the classical limit. These
are the FM and stripy AF states {[}shown as right panels in Fig.~\ref{fig3}{]}
with spin ordering along the $z$ axis. Within spin-wave theory we
find one branch in the FM state 
\begin{align}
 & \omega^{2}\left({\bf q}\right)=\left(K_{z}-K_{x}c_{x}\right)\left(K_{z}-K_{y}c_{y}\right)~,\label{eq4}
\end{align}
and four branches in the stripy AF state 
\begin{align}
 & \omega_{1,2}^{2}\left({\bf q}\right)=K_{z}^{2}\pm\sqrt{K_{z}^{2}\left(K_{x}^{2}c_{x}^{2}+K_{y}^{2}s_{y}^{2}\right)-K_{x}^{2}K_{y}^{2}c_{x}^{2}s_{y}^{2}}~,\label{eq5}
\end{align}
where $c_{x}=\cos\mathbf{q}\cdot\mathbf{a}_{x}$, $c_{y}=\cos\mathbf{q}\cdot\mathbf{a}_{y}$,
$s_{x}=\sin\mathbf{q}\cdot\mathbf{a}_{x}$ and $s_{y}=\sin\mathbf{q}\cdot\mathbf{a}_{y}$.
The other two branches, $\omega_{3,4}({\bf q})$, are obtained from
$\omega_{1,2}\left({\bf q}\right)$ by the exchanges $c_{x}\leftrightarrow c_{y}$
and $s_{x}\leftrightarrow s_{y}$.

Fig.~\ref{fig3} shows the color map of the obtained spin-wave excitation
spectra for $K_{\gamma}=1$ in both the FM state and the stripy AF
state. In this latter case, only one of the two degenerate lower branches
is reported. The corresponding magnetic structures are sketched on
the right. In both cases, the excitation spectra are well defined
over the entire Brillouin zone, confirming that the FM and the stripy AF states
do indeed minimise the classical energy. Moreover, the lines of nodes, related to the sub-extensive
degeneracy of the classical manifold discussed above, are clearly visible.
Comparing the zero-point spin-wave energies obtained from Eqs.~(\ref{eq4})
and (\ref{eq5}), we find that the FM state is always favored against
the stripy AF state, in agreement with the linked-cluster expansion
result. Moreover, by expanding the spin-wave excitation spectra in
Eqs.~\ref{eq4} and \ref{eq5} in terms of small $K_{x,y}/K_{z}$,
we find analytically the difference between the zero-point energies
of the FM and the stripy AFM states to be $\delta E_{{\rm sw}}^{(4)}=-\frac{3}{4}\times\frac{1}{192}\frac{K_{x}^{2}K_{y}^{2}}{K_{z}^{3}}$,
which is in agreement with the prediction of Eq.~(\ref{eq3}) except
for a multiplicative factor $\frac{3}{4}$, whose presence can be
anyway readily explained. The linear spin-wave theory does not take
into account the interactions between misaligned spins. Therefore,
in the virtual state shown as the middle \emph{diamond} in Fig.~\ref{fig1}(c),
8 broken $\left(z\right)$-bonds are counted in the linear spin-wave
theory instead of the actual 6 broken $\left(z\right)$-bonds shown
in Fig.~\ref{fig1}(c).

In conclusion, we have discussed quantum order-by-disorder in the
Kitaev model on the triangular lattice within the linked-cluster expansion
and the complementary spin-wave theory, and clarified the true nature
of the ground state of this frustrated quantum spin model. In particular,
we have shown (i) the presence of a mechanism of quantum selection
of easy axes, (ii) the emergence of a four-spin interaction
that reduces the sub-extensive degeneracy of the ground state manifold
($3\times2^{L}$) down to a discrete one ($3\times2^{2}$), and (iii)
the existence of a hidden symmetry of the model that protects this
latter degeneracy. The present analytical study explains and quantifies
the results of numerical simulations \cite{Bec14}. The developed
scheme, that makes explicit links between degenerate perturbation
theory and spin-wave analysis, can be applied to other quantum spin
models in which spin frustration is driven by anisotropic spin couplings.

We thank A.F.~Bangura, M.~Becker, M.~Daghofer, M.~Hermanns, G.~Khaliullin, 
 S.~Trebst and E.A.~Yelland for discussions and comments.
AA acknowledges kind hospitality at the Max Planck Institute for Solid State Research, Stuttgart, Germany.

\bibliographystyle{apsrev4-1}

%

\end{document}